\documentclass{ws-ijmpb}

\begin{document}

\def \ba {\begin{eqnarray}}
\def \ea {\end{eqnarray}}

\markboth{Choi and Bok}
{How to pin down the pairing interaction for high $T_c$ superconductivity in cuprates}

%
\catchline{}{}{}{}{}
%

\title{How to pin down the pairing interaction for high $T_c$ superconductivity in cuprates }

\author{Han-Yong Choi}

\address{Department of Physics and Institute for Basic Science Research, Sungkyunkwan University, Suwon 16419, Korea\\
hychoi@skku.edu}

\author{Jin Mo Bok}

\address{Department of Physics and Institute for Basic Science Research, Sungkyunkwan University, Suwon 16419, Korea\\
bockill@skku.edu}

\maketitle

\begin{history}
\received{Day Month Year}
\revised{Day Month Year}
\end{history}

\begin{abstract}

The normal and pairing self-energies are the microscopic quantities which reflect and characterize the underlying interaction in superconductors. The momentum and frequency dependence of the self-energies, therefore, provides the experimental criteria which can single out the long sought-after pairing interaction among many proposed ideas. This line of research to pin down the pairing interaction for the cuprate superconductors has been carried out with some success by analyzing the momentum distribution curves of laser angle-resolved photo-emission spectroscopy (ARPES) data. Some progress and results are presented and compared with theoretical calculations based on leading proposals. Comments are made on the proposed scenarios from the comparisons.

\end{abstract}


\section{Introduction}

Impressive progresses have been made since the discovery of the cuprate high $T_c$ superconductors more than thirty years ago.\cite{1} They range from materials quality, new and refined experimental tools to novel theoretical ideas and techniques. Despite enormous new findings along with plethora of phases, some of key questions, like the pairing mechanism of the high $T_c$ superconductivity, have not been settled down yet.\cite{2} This is not due to a lack of ideas. Rather it is because we do not have clear and concrete criteria to differentiate among proposed ideas. We wish to suggest such criteria in terms of the quasi-particle self-energy. We present and review some progresses made along this line from the self-energy analysis of laser ARPES intensity data of the bilayer Bi$_2$Sr$_2$CaCu$_2$O$_{8+\delta}$ (Bi2212) samples.\cite{3}

There are four constraints any viable pairing theory for the cuprates must meet. The first and obvious constraint, the $d$-wave pairing having been firmly established for the cuprates, is if a proposed theory gives rise to a superconducting (SC) gap of $d$-wave symmetry. In terms of the self-energy, this corresponds to (1) $\theta$ dependence of the pairing self-energy $\phi(\theta,\omega)$ having $d$-wave angle dependence like $\phi(\theta,\omega) \sim \cos(2\theta)$. The angle $\theta$ refers to the direction of the inplane wavevector ${\bf k}$ with respect to the crystalline $x$-axis. The others are (2) $\omega$ dependence of $\phi(\theta,\omega)$, $\omega$ being the energy of conduction electrons with respect to the chemical potential. (3) $\theta$ dependence of the normal self-energy $\Sigma(\theta,\omega)$, and (4) $\omega$ dependence of $\Sigma(\theta,\omega)$. To wrap up, our proposal to pin down the underlying pairing interaction for the cuprates is to check if a proposed theory produces the $\theta$ and $\omega$ of the normal and pairing self-energies such that they match those extracted from experiments. We provide such experimentally determined self-energies from laser ARPES from Bi2212 samples here.

The self-energy analysis makes fittings to the momentum distribution curve (MDC) using the superconducting Green's function and extract the normal and pairing self-energies as a function of the angle $\theta$ and frequency $\omega$ at several temperatures above and below $T_c$.\cite{4}  All the results presented here were obtained by analyzing laser-based ultra high resolution ARPES data for slightly underdoped Bi2212 with $T_{c}=89$ K (UD89) and overdoped with $T_{c}=82$ K (OD82) samples. Both are measured in highly stable vacuum condition with $\sim$ 1 meV energy and $\sim$ 0.004 {\AA}$^{-1}$ momentum resolution.

This line of thoughts is nothing new actually.\cite{5,6} The celebrated McMillan-Rowell approach which firmly established the phonon-mediated $s$-wave pairing for the conventional superconductors is indeed to check the $\omega$ dependence of the SC gap $\Delta(\omega)$. See Eqs.\ $(1)-(3)$ below for formal relations between the gap and self-energies. The $\omega$ dependence of $\Delta(\omega)$ of lead was determined by measuring the tunneling conductance $dI/dV$ and was inverted to calculate the Eliashberg function $\alpha^2 F(\omega)$. It was compared with the phonon density of states measured from the inelastic neutron scattering  experiments. A very good agreement between the $\alpha^2 F(\omega)$ obtained by the Eliashberg inversion and the phonon density of states measured by the neutron scattering firmly established the phonon mediated pairing of superconductivity.\cite{5} A main difficulty of applying the McMillan-Roewll analysis to the cuprates is the $d$-waveness of the SC pairing.\cite{6} The normal self-energy has the full lattice symmetry but the pairing self-energy has the $d$-wave symmetry. One must resolve the momentum as well as the frequency dependence in the analysis. The ARPES can provide such experimental data. The details of the analysis for $d$-wave pairing was published before.\cite{3,4}

\section{Extracted self-energies}

One of main results of the self-energy analysis is presented in Fig.\ 1. In Fig.\ 1(A), we show the real ($\Sigma_1$) and imaginary ($\Sigma_2$) parts of the extracted normal self-energy as a function of the frequency $\omega$ along the cut of $\theta=20^\circ $ from Bi2212 OD82 sample at five temperatures above and below $T_c$. The normal self-energy at $T=90$ K (above $T_c$) is in accord with the established results. The imaginary part exhibits the anticipated marginal Fermi liquid behavior; being linear in $\omega$ for $T<|\omega|<\omega_c $.\cite{7} The real part is also in agreement with previous results. It approximately satisfies the Kramers-Kronig relation from the imaginary part and makes the zero crossing around $\approx-0.13$ eV. In SC state, $\Sigma(\theta,\omega)$ exhibits two energy scales around $\omega \approx -50$ meV and $-10$ meV. One around $-50$ meV was expected but the other was not. A peak energy in the self-energy is given by $\omega_p \approx \omega_b +\Delta$, $\omega_b$ is the peak energy of the mediating boson. For electronically mediated pairing the boson fluctuation peak is around $2\Delta$ in the SC state. Therefore a peak around $\omega_p \approx 3\Delta$ was expected and indeed appeared around 50 meV. This energy accidentally overlaps with the broad peak energy position as observed in the dispersion kink by ARPES in the normal as well as SC states.\cite{8} The low energy peak around 10 meV $<\Delta$, on the other hand, was unexpected. It was understood as being induced by the forward scattering impurities located in between CuO$_2$ planes.\cite{9} The low energy dispersion kinks reported previously from ARPES experiments were explained in terms of this forward scattering impurity peak.\cite{10}

In Fig.\ 1(B) the real and imaginary parts of the pairing self-energy are presented for three cuts of $\theta=$ 20, 25, 30$^\circ$ from OD89 sample deep in the SC state at $T=16$ K. This being the first determination of the pairing self-energy, perhaps some arguments for its reliability may be necessary. What are plotted is the scaled pairing self-energy $\phi(\theta,\omega)/\cos(2\theta)$. For a simple $d$-wave BCS case, they will collapse onto a single curve. The fact that the three curves collapse reasonably well onto a single curve means that the extracted pairing self-energy is in accord with the $d$-wave pairing. The pairing self-energy also exhibits two energy scales as the normal self-energy from OD82; one around $3\Delta \approx 60$ meV and another impurity peak $\approx 10$ meV. $\Delta$ from UD89 is slightly larger than OD82 because of higher $T_c$. The off plane impurity features may be analyzed to obtain the impurity parameters. The fact that the impurity scattering parameters from the normal and pairing self-energies agree well with each other gives another support that the pairing self-energy is extracted properly. See Table 1 below. One technical comment is that it is important to take the particle-hole asymmetricity into consideration to successfully extract the impurity feature in the self-energies, that is, include the dispersion shift $X $ function in the SC Green's function. See equation (2) below.

\begin{figure}
\center
\vspace*{8pt}
\includegraphics[scale=0.45]{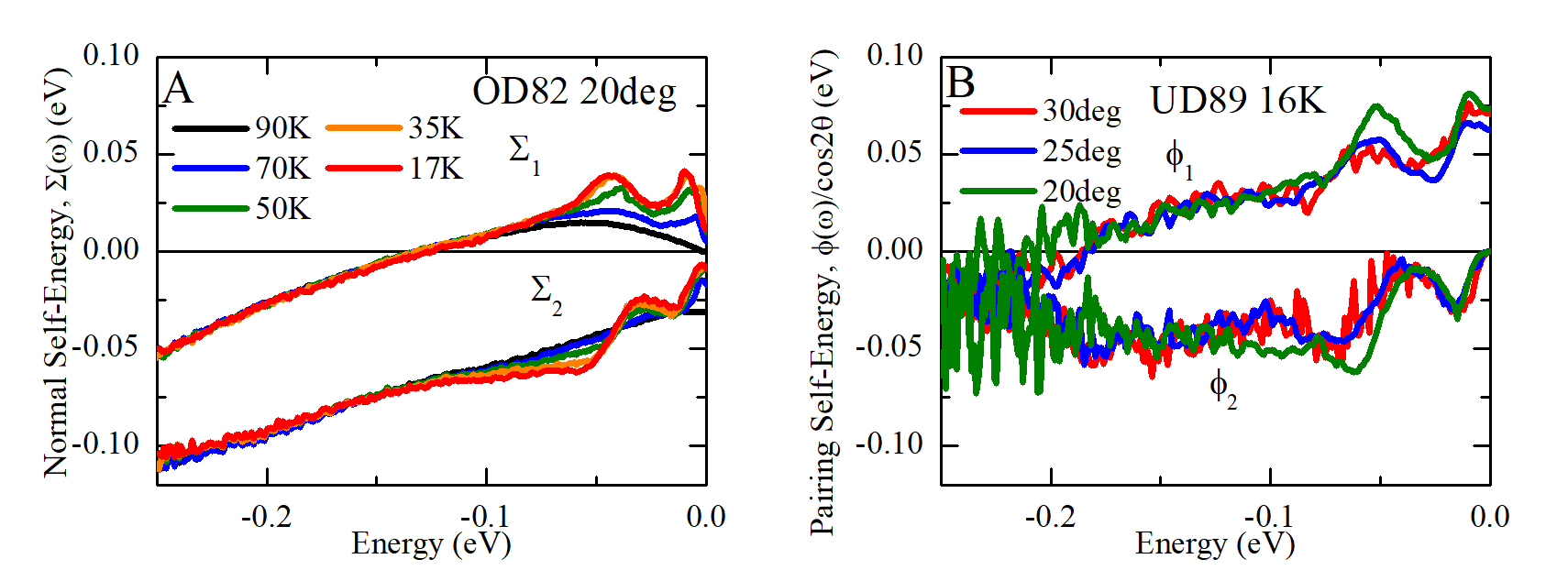}
\caption{The extracted self-energies from Bi2212 OD82 and UD89 samples. (A) The normal self-energy along $\theta=20^\circ$ cut at five different temperatures above and below $T_c$ from OD82 sample. The SC induced features emerge at two energy positions around $\approx -50$ and $-10$ meV. (B) The scaled pairing self-energy $\phi(\theta,\omega)/\cos(2\theta)$ as a function of $\omega$ along $\theta=20$, 25, and 30 degree cuts from UD89 sample. The scaling exhibits the consistency with the $d$-wave pairing. There also emerge two SC induced energy scales.}
\end{figure}

\section{Forward scattering impurity and Green's function}

The ARPES intensity is given by the spectral function which is the imaginary part of the retarded Green's function. The Green's function is given by

 \ba \label{greens}
G(\mathbf{k},\omega)=
\frac{W(\mathbf{k},\omega)+Y(\mathbf{k},\omega)}
{W^{2}-Y^{2}-\phi^{2}},
 \ea
where
 \ba
W(\mathbf{k},\omega)=
\omega-\widetilde{\Sigma}(\mathbf{k},\omega) = \omega Z(\mathbf{k},\omega),~~
Y(\mathbf{k},\omega)= \xi(\mathbf{k})+X(\mathbf{k},\omega),
 \ea
$\xi(\mathbf{k})$ is the quasi-particle bare dispersion,
$X(\mathbf{k},\omega)$ is the dispersion shift,
and 
 \ba
\Sigma(\mathbf{k},\omega) =
\widetilde{\Sigma}(\mathbf{k},\omega) + X(\mathbf{k},\omega)
 \ea
is the normal self-energy.\cite{11,4} The gap function $\Delta$ is given by
 \ba
\Delta(\mathbf{k},\omega) =\phi(\mathbf{k},\omega)/Z(\mathbf{k},\omega).
 \ea

The impurity scattering may be described as follows. The self-energy consists of two terms due to the coupling to a boson spectrum and to impurities.
 \ba
\Sigma(\mathbf{k},\omega)=\Sigma_{eff}(\mathbf{k},\omega)+\Sigma_{imp}(\mathbf{k},\omega), \\
\phi(\mathbf{k},\omega)=\phi_{eff}(\mathbf{k},\omega)+\phi_{imp}(\mathbf{k},\omega).
 \ea
The subscript $eff$ represents the self-energy induced by boson spectrum. The impurity term can be written as
\begin{eqnarray}
\Sigma_{imp}(\mathbf{k},\omega)=
n_{imp}\sum_{\mathbf{k}'}|V_{imp}(\mathbf{k},\mathbf{k}')|^{2}\frac{W(\mathbf{k},\omega)
+Y(\mathbf{k},\omega)}{W^{2}(\mathbf{k},\omega)-Y^{2}(\mathbf{k},\omega)
-\phi^{2}(\mathbf{k},\omega)}, \label{sigmaimp}
\end{eqnarray}
where the $n_{imp}$ and $V_{imp}$ are the impurity concentration and impurity potential, respectively. $\phi_{imp}$ may also be written similarly.
\begin{equation}
\phi_{imp}(\mathbf{k},\omega)=-n_{imp}\sum_{\mathbf{k}'}|V_{imp}(\mathbf{k},\mathbf{k}')|^{2}\frac{\phi(\mathbf{k},\omega)}{W^{2}(\mathbf{k},\omega)-Y^{2}(\mathbf{k},\omega)-\phi^{2}(\mathbf{k},\omega)}, \label{phiimp}
\end{equation}

A reasonable model for impurity potential was suggested before,\cite{9}
\begin{equation}
V_{imp}(\mathbf{k},\mathbf{k}')=\frac{2\pi\kappa V_{0}}{[(\mathbf{k}-\mathbf{k}')^{2}+\kappa^{2}]^{3/2}},
\end{equation}
where $\kappa$ is the range of the impurity potential that controls the angle dependence of the impurity self-energy. The normal and pairing self-energies including this impurity potential scattering may be written as
\begin{eqnarray}
\Sigma(\mathbf{k},\omega) &=& \Sigma_{imp}(\mathbf{k},\omega)+i\Gamma,  \nonumber  \\
\phi(\mathbf{k},\omega) &=& \phi_{imp}(\mathbf{k},\omega)+\phi_{0}[\cos(k_{x}a)-\cos(k_{y}a)]/2, \label{impeq}
\end{eqnarray}
where the frequency dependence of $\Gamma$ and $\phi_0$ was neglected because the relevant frequency range is limited for impurity scattering.
The impurity scattering parameters were determined by matching the extracted self-energies. Eq.\ (\ref{impeq}) was solved self-consistently together with Eqs.\ (\ref{sigmaimp}) and (\ref{phiimp}). The determined parameters are presented in Table 1. The low energy structure can be eliminated by model impurity self-energy as shown in Fig.\ 2. The determined parameters are quite similar for normal and pairing impurity self-energies as alluded above.

\begin{table}[ph]   
\tbl{Parameters of the impurity scattering potential from fitting the imaginary part of the extracted normal and pairing self-energies.}
{\begin{tabular}{lllll} \Hline
\\[-1.8ex]
Parameter & $\Sigma_{2}$ & {} & $\phi_{2}$ & {}\\[0.8ex]
{} & 25deg & 20deg & 25deg & 20deg \\
\hline \\[-1.8ex]
$\kappa (1/a)$& 0.3 & 0.3  & 0.3 & 0.3\\
$n_{imp}V_{0}^{2} (t^{2}a\kappa^{3})$ & 0.03 & 0.03  & 0.035 & 0.03\\
$\Gamma_{0}$ (meV) & 3.5 & 3.5 & 5 & 3 \\
$\phi_{0}$ (meV) & 18 & 18 & 20 & 18 \\ [0.8ex]
\Hline \\[-1.8ex]
\end{tabular}}
\label{tab1}
\end{table}

\begin{figure}
\center
\vspace*{8pt}
\includegraphics[scale=0.4]{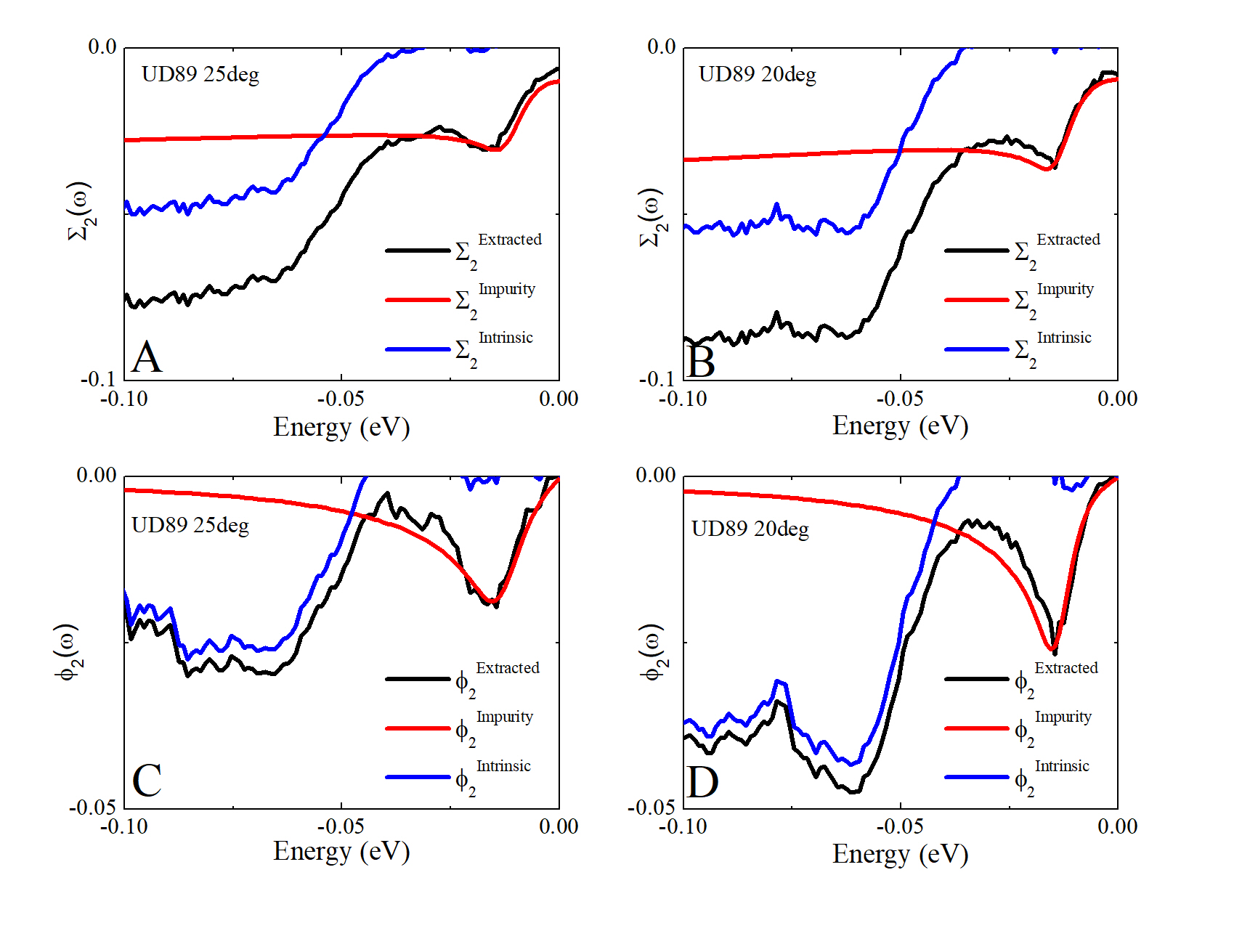}
\caption{The imaginary part of normal and pairing self-energies along $\theta=20$ and 25 degree cuts. Black lines are extracted self-energies from UD89 sample data. Red lines are calculated self-energy induced by forward scattering impurities. Blue lines are intrinsic self-energy after subtracting the impurity self-energy from the extracted one. The fitted impurity scattering parameters are list in the Table 1. }
\end{figure}

\section{Intrinsic pairing and normal self-energies}

The impurity feature may serve as a flag which signals a successful extraction of self-energy. This feature, however, being from the extrinsic impurity scatterings, must be eliminated to give the intrinsic self-energy. This procedure was explained in the previous section. The obtained intrinsic self-energies and the gap function from Bi2212 UD89 are shown in Fig.\ 3. The experimental error bars in the ARPES intensity data do not allow us to go beyond $\sim -0.15$ eV to reliably extract the pairing self-energy. They exhibit a single energy scale around $\omega_p \approx 3\Delta \approx 60$ meV. $\Delta\approx 20 $ meV agrees with the $\Delta$ from the off-plane impurity scattering parameters as listed in Table 1.

In some cases it is easier to compare the Eliashberg functions rather than the self-energies. The self-energy can be obtained without a specific theory. The only relation one needs is that the ARPES intensity is given by the spectral function. But, to obtain the Eliashberg function from the extracted self-energy one needs a theory relating them. It is the Eliashberg equation which relates the normal self-energy with the normal Eliashberg function $\alpha^2 F_n (\theta,\omega)$, and the pairing self-energy with the pairing Eliashberg function $\alpha^2 F_p (\theta,\omega)$. The calculated Eliashberg functions from the intrinsic self-energies were presented before.\cite{3}
But there are some issues about validity of the Eliashberg equation for the cuprtes. So it will be best if one can complete the check using the self-energies without involving the Eliashberg functions.

\begin{figure}
\center
\vspace*{8pt}
\includegraphics[scale=0.4]{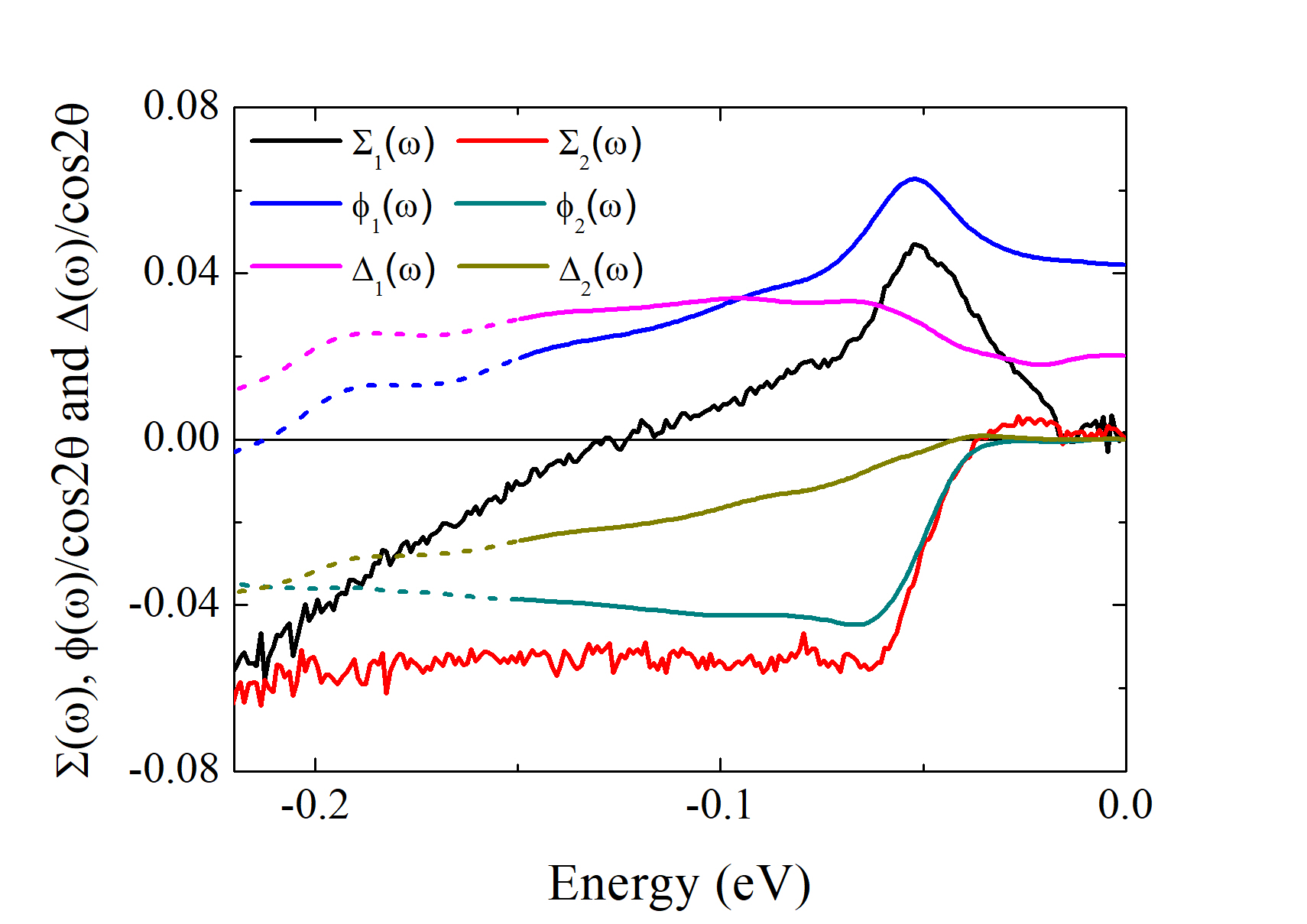}
\caption{The intrinsic normal and scaled pairing self-energies and gap function along $\theta=20\deg$ from UD89 sample after the impurity subtraction shown in Fig.\ 2. Only single SC induced energy scale shows up around $\approx 3\Delta$. }
\end{figure}

\section{Comparison of leading proposals against experimental self-energies}

We now try to check some proposed ideas against above extracted results. Let us consider here three leading ideas among many proposals.\cite{2,12} (a) Antiferromagnetic (AF) spin fluctuation theory,\cite{13} (2) resonating valence bond (RVB) theory,\cite{14} and (3) loop current fluctuation theory.\cite{15} What must be done is to calculate the normal and pairing self-energies as a function of $\theta$ and $\omega$ based on each proposal both above and below $T_c$ and compare them with the extracted results in Fig.\ 3.

\begin{figure}[t]
\center
\vspace*{8pt}
\includegraphics[scale=0.4]{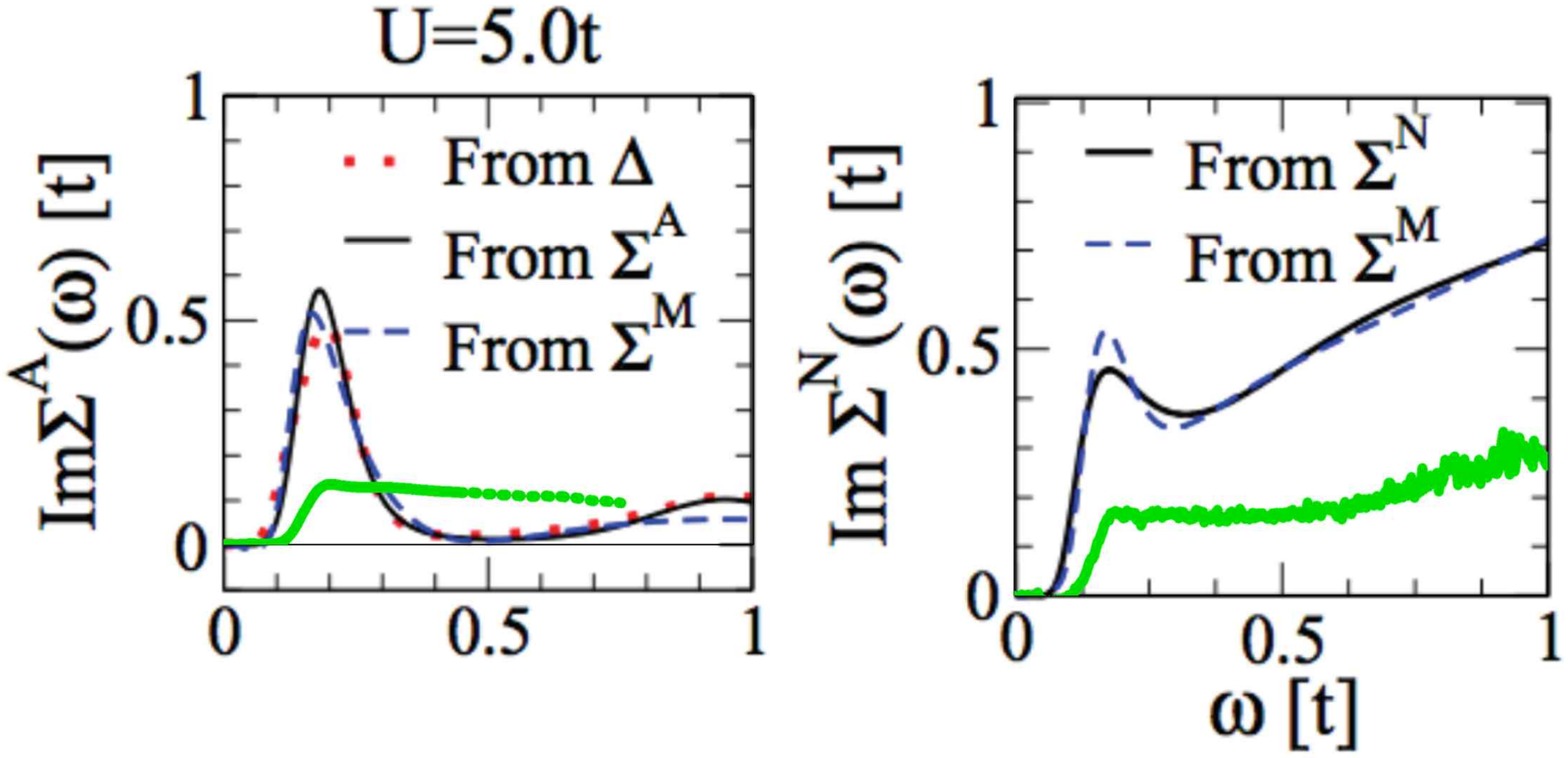}
\vspace{-1.5cm}
\caption{A comparison of the intrinsic self-energies with the cluster-DMFT calculations of the two-dimensional Hubbard model.$^{19}$[reproduced under permission] The case of $U=5.0 t$ is shown here. The thick green curves are the extracted intrinsic self-energies, and different lines of DMFT calculations represent different methods to obtain the self-energies. The left pannel compares the imaginary part pairing self-energy and the right panel the normal self-energy. Notice the different peak weights around $\approx 0.2 t \approx 60$ meV. }
\end{figure}

\subsection{AF spin fluctuation theory}

The AF spin fluctuation theory was checked previously along this line. The angle dependence of the extracted $\alpha^2 F_n (\theta,\omega)$ was compared with the phenomenological AF spin spectrum. The normal state $\alpha^2 F_n (\theta,\omega)$ obtained by inverting normal self-energy from UD89 sample exhibits almost momentum isotropic behavior except for the angle dependent cutoff energy.\cite{16} This is at odds with the AF spin spectrum. The AF spin fluctuations must have strong enough momentum anisotropy to induce high $T_c$ $d$-wave pairing. Because the spin fluctuation theory has a single scala vertex to couple to electrons, the similar momentum anisotropy must appear in both normal and pairing channels for a generic AF spectrum. The difference comes from the projection of the momentum structure of the AF spin fluctuations onto the $s$-wave channel for $\alpha^2 F_n (\theta,\omega)$ and onto the $d$-wave channel for $\alpha^2 F_p (\theta,\omega)$.\cite{16}

This check was also done in SC state in terms of the self-energies. The experimental spin spectra reported by inelastic neutron scattering from optimally doped  {La$_{2-x}$Sr$_x$CuO$_4$} samples\cite{17} were adopted. The full momentum and frequency dependent spectrum was used in the Eliashberg equation with the realistic tight-binding electron dispersion to calculate the normal and pairing self-energies. The calculated $\theta$ and $\omega$ dependence of the self-energies, as might be expected from the normal state results, was again at odds with the extracted self-energies.\cite{18}

Also, the cluster dynamical mean-field theory (DMFT) calculations were performed to solve the two-dimensional Hubbard model to obtain the normal and pairing self-energies.\cite{19} This should be free from a criticism  of the Eliashberg equation to describe the cuprates. For appropriate range of $U/t$, the results may be relevant for AF fluctuation theory, and for a large value, should be relevant for RVB theory description. But, $U$ values larger than the bandwidth are not computationally accessible at present, and the number of sites in the cluster is not enough to resolve momentum dependence of the self-energies. Yet, a conspicuous feature of the DMFT calculations is a strong peak around $\omega\approx 0.2 t \approx$ 60 meV. The extracted self-energy only exhibits a smaller jump at the same energy. Also DMFT results for the pairing self-energy shows a suppression around $\approx 0.5 t$ while that does not appear in the extracted one. Compare the imaginary parts of the self-energy in Fig.\ 4. The frequency dependence of the normal and pairing self-energies from the Hubbard model DMFT calculations do not agree with the extracted counter-parts. This discrepancy gets enhanced as $U$ is increased.

\subsection{RVB theory}

How does the RVB theory compare against the experimental self-energy? As mentioned above, the DMFT calculation can not yet access the parameter regime thought to be relevant to RVB physics. It is amusing that despite so enormous research in the RVB camp, the frequency dependent self-energies from RVB theory are not available yet to be checked against the extracted self-energies from experiments.

An extension of the ARPES data in terms of the frequency range and resolution will help settle down the question about the pairing glue.\cite{20,21} The pairing self-energy can be written using the Kramers-Kronig relation,
 \begin{equation}
\phi_{1}(\theta,\omega=0)=\frac{2}{\pi}\int_0^\infty d\omega'
 \frac{\phi_2(\theta,\omega')}{\omega'} +\phi_{NR}.
 \end{equation}
Here, $\phi_{NR}$ is frequency independent (non-retarded) part of pairing self-energy. This corresponds to the instantaneous pairing which does not involve the pairing glue.\cite{21}  A small value of the ratio $\phi_{NR}/\phi_{1}(\theta,\omega=0)$ compared with 1 means that the pairing glue is a very much valid concept, while the ratio $\approx 1$ implies there is no need for pairing glue. Currently $\phi_{1}(\theta,\omega=0)$ can be determined accurately but the integral of  $\phi_2(\theta,\omega)/\omega$ over $\omega$ has some uncertainty because $\phi_2(\theta,\omega)$ can only be extracted upto $\omega \approx 0.15$ eV. With improvement of the frequency range by about a factor of 2 for the ARPES data, which is expected to go beyond the cutoff energy, we can determine the ratio with more accuracy. This will shed light on the question about validity of pairing glue for cuprates.

\subsection{Loop current theory}

Now comes Varma's loop current theory.\cite{15} Here again, the $\theta$ and $\omega$ dependent self-energies from loop current fluctuation theory are not available yet to be checked. This is less surprising because the loop current description needs three band Hubbard model which is computationally much more expensive. Nevertheless, one can try comparison between Eliashberg functions.

Recall the dilemma of the AF spin fluctuation theory exposed above: the normal Eliashberg function $\alpha^2 F_n (\theta,\omega)$ must be momentum anisotropic in AF theory to be consistent with the high $T_c$, in contradiction to the experimentally determined $\alpha^2 F_n (\theta,\omega)$. Or, put differently, ``the central paradox'': The same fluctuations which is strong enough in the pairing channel must lead to a nearly angle-independent scattering in the normal channel.\cite{3} How can this be brought about? The essence of Varma idea is that the loop current fluctuations, interestingly, have the vector vertex to couple to the electrons. This yields two distinct couplings 1 and $\cos(2\theta)\cos(2\theta')$ along the normal and pairing channels, respectively. The consequence is that
 \ba
\alpha^2 F_n (\theta,\omega) = c ~\alpha^2 F_p (\theta,\omega)/\cos(2\theta),
  \ea
where $c$ is a constant of order 1. With this statement our results over the available frequency range seem to be consistent.\cite{3} It will of course be desirable if we can make the comparison over wider frequency range which covers the angle dependent cutoff energy.

\section{Summary and Outlooks}

An important message of the this work is that the intrinsic normal and pairing self-energies can indeed be obtained from experiments. Their angle and frequency dependence provides the most stringent check for the underlying pairing interaction for the cuprate high $T_c$ superonductors. We have presented and reviewed some of the self-energies extracted from the ultra high resolution laser ARPES intensity data from Bi2212 samples. One of remaining issues in this direction is to extend the reliable energy range in measured ARPES data. The next generation ARPES setup like the angle resolved time-of-flight (ARToF) detector combined with 11 eV laser seems promising in this direction.

\section*{Acknowledgements}

We would like to thank Chandra Varma and Xingjiang Zhou for the collaborations and conversations over the years. This work was supported by Samsung Science \& Technology Foundation through the Grant No.\ SSTF-BA1502-06.



\section*{References}

\begin{thebibliography}{0}


\bibitem{1} B. Keimer, S. A. Kivelson, M. R. Norman, S. Uchida, and J. Zaanen, Nature \textbf{518}, 179 (2015).

\bibitem{2} E. Abrahams, Int. J. Mod. Phys. B {\bf 24}, 4150 (2010).

\bibitem{3} J. M. Bok {\it et al.}, Science Adv. \textbf{2}, e1501329 (2016).
	
\bibitem{4} J. H. Yun {\it et al.}, Phys. Rev. B {\bf 84}, 104521 (2011).

\bibitem{5} W. L. McMillan and J. M. Rowell, Phys. Rev. Lett. {\bf 14}, 108 (1965).

\bibitem{6} I. Vekhter and  C. M. Varma, Phys. Rev. Lett. {\bf 90}, 237003 (2003).

\bibitem{7}  C. M. Varma, P. B. Littlewood, S. Schmitt-Rink, E. Abrahams, and A. E. Ruckenstein, Phys. Rev. Lett. \textbf{63} 1996 (1989).

\bibitem{8} J. He {\it et al.}, Phys. Rev. Lett. {\bf 111}, 107005 (2013).

\bibitem{9} L. Zhu, P. J. Hirschfeld, and D. J. Scalapino, Phys. Rev. B {\bf 70}, 214503 (2004).

\bibitem{10} S. H. Hong {\it et al.}, Phys. Rev. Lett. {\bf 113}, 057001 (2014).
	
\bibitem{11}  A. W. Sandvik, D. J. Scalapino, and N. E. Bickers, Phys. Rev. {\bf 69}, 094523 (2004).

\bibitem{12} H.-Y. Choi, J. Kor. Phys. Soc. {\bf 60}, 978 (2012).

\bibitem{13} D. J. Scalapino, Rev. Mod. Phys. {\bf 84},1383 (2012).

\bibitem{14} P. W. Anderson, Science {\bf 235}, 1196 (1987).

\bibitem{15} C. M. Varma, Phys. Rev. B {\bf 73}, 155113 (2006).

\bibitem{16} J. M. Bok {\it et al.}, Phys. Rev. B {\bf 81}, 174516 (2010).

\bibitem{17} B. Vignolle {\it et al.}, Nature Phys. {\bf 3}, 163 (2007).

\bibitem{18} S. H. Hong and H. Y. Choi, J. Phys.: Condensed Matt. {\textbf 25}, 365702 (2013).

\bibitem{19} E. Gull and A. J. Millis, Phys. Rev. B {\bf 91}, 085116 (2015).

\bibitem{20} P. W. Anderson, Science {\bf 316}, 1705 (2007).

\bibitem{21} T. A. Maier, D. Poilblanc, and D. J. Scalapino, Phys. Rev. Lett. {\bf 100}, 237001 (2008).


\end{thebibliography}

\end{document}